\documentclass[aps,prl,10pt,notitlepage,noeprint,twocolumn,floatfix,longbibliography]{revtex4-2}
\usepackage{amsmath,amssymb,amsfonts,bm,comment}
\usepackage{graphics,float,xcolor}
\usepackage[caption=false]{subfig}
\usepackage[colorlinks=true,
    linkcolor=blue,
    filecolor=magenta,
    urlcolor=blue,
    citecolor=blue]{hyperref}
\usepackage{physics,xparse,mathtools,isotope,soul}
\usepackage{lipsum}
\usepackage[note-name = ,use-sort-key = false]{notes2bib} 
\usepackage{microtype}

\usepackage[table-parse-only]{siunitx}
\newcommand{\im}{\mathrm{i}}

\newcommand{\figref}[1]{\unskip\;\textup{\ref{#1}}}

\makeatletter
\let\origref\ref
\renewcommand{\eqref}[1]{\unskip\;\textup{\tagform@{\origref{#1}}}}
\makeatother

\begin{document}
\title{Unveiling Supersolid Order via Vortex Trajectory Correlations}

\author{Subrata Das}
\email{subrata@vt.edu}
\author{Vito W. Scarola}
\email{scarola@vt.edu}
\affiliation{Department of Physics, Virginia Tech, Blacksburg, Virginia 24061, USA}
\date{\today}

\begin{abstract}
    The task of experimentally investigating the inherently dual properties of a supersolid, a simultaneous superfluid and solid, has become more critical following the recent experimental exploration of supersolid regimes in dipolar Bose-Einstein condensates (BECs) of $^{164}\text{Dy}$.  We introduce a supersolid order parameter that uses vortex-vortex trajectory correlations to simultaneously reveal the periodic density of the underlying solid and superfluidity in a single measure.  We propose experiments using existing technology to optically create and image trajectories of vortex dipoles in dipolar BECs that is applicable to large system sizes.
    We numerically test our observable and find that vortex-vortex correlations reveal the supersolid lattice structure while distinguishing it from superfluidity even in the presence of dissipation.   Our method sets the stage for experiments to use vortex trajectory correlations to investigate fundamental properties of supersolids arising from their dynamics and phase transitions as experimental system sizes are increased.
\end{abstract}

\maketitle

\begin{figure}[t]
    \centering
    \includegraphics[width=0.49\textwidth]{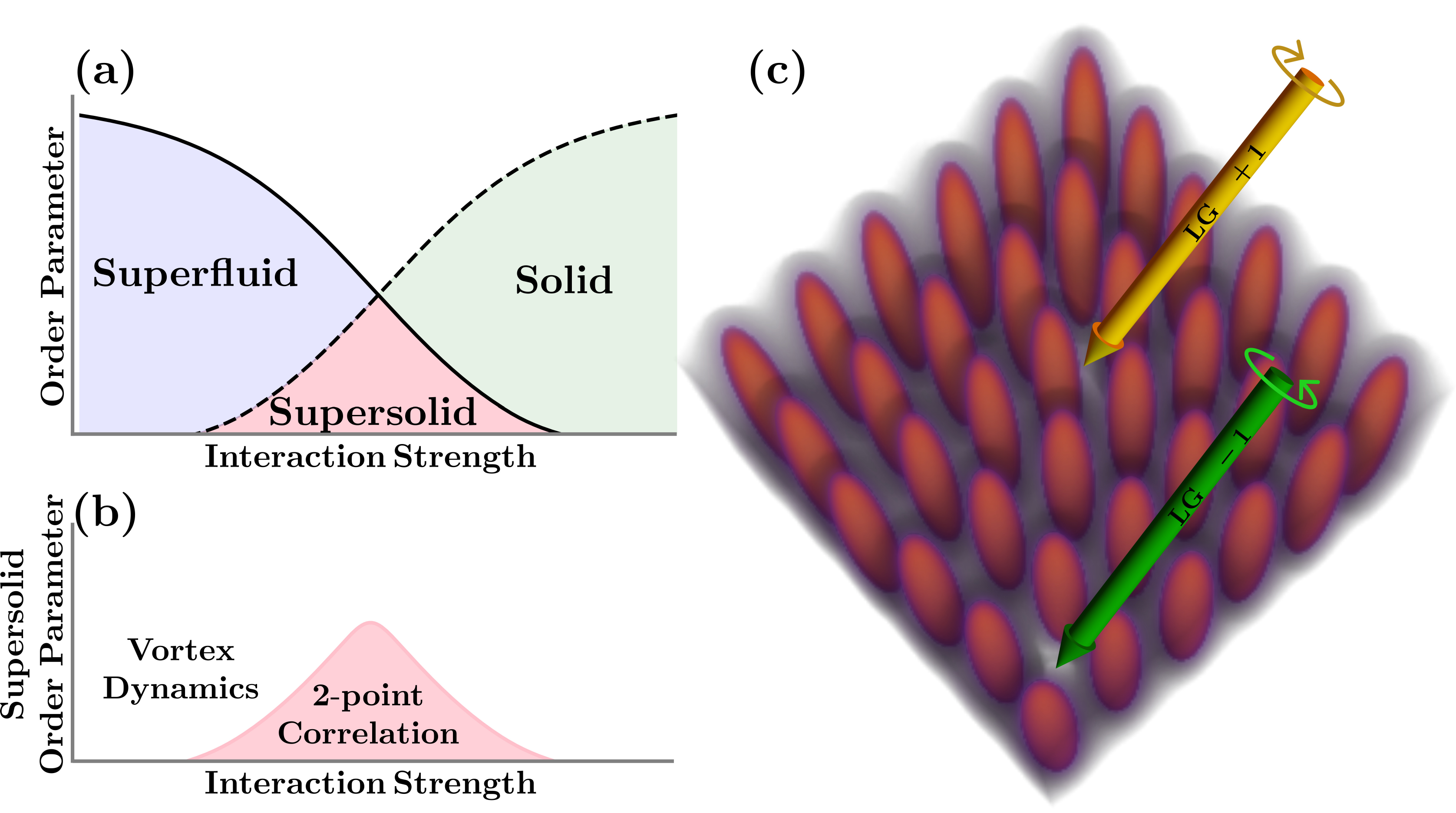}
    \caption{(a) Schematic phase diagram where weak particle-particle interaction (compared to kinetic energy) favors a superfluid and strong interaction favors a solid.  Overlapping order parameters define a supersolid.  (b) The same as (a), but here, a nonzero vortex-vortex trajectory correlation function offers a distinct supersolid order parameter.  (c) Proposal to create a vortex dipole amid quantum droplets defining a quasi-two-dimensional supersolid in a dipolar BEC using a pair of Laguerre-Gaussian beams. }
    \label{fig:schema_SS_LG_combo}
\end{figure}

The coexistence of a superfluid and solid at the same time and place defines a supersolid \cite{gross_1957_unified,andreev_1969_quantum,chester_1970_speculations,boninsegni_2012_colloquium}.  Superfluidity arises as a physical manifestation of off diagonal long-range order (ODLRO) in the density matrix triggered by the spontaneously broken $U(1)$ gauge symmetry of the many-body wave function.  The presence of metastable quantized vortices offers a sufficient condition for identification of ODLRO \cite{leggett_2006_quantum}.  Whereas the solid component arises from diagonal long-range order (DLRO) in the density matrix derived from spontaneously broken translational symmetry, signatures of DLRO include long-range spatial oscillations in particle-particle pair correlations.  Combined observation of such overlapping long-range order parameters signals a macroscopic quantum state, a supersolid, Fig. \figref{fig:schema_SS_LG_combo}(a).

Supersolids are challenging to study with globally averaged observables.  Experiments \cite{kim_2004_probable} proposed \cite{leggett_1970_can} to reveal nonclassical moments of inertia of supersolid $^4\text{He}$ were brought into question \cite{kim_2012_absence,chan_2013_overview} by the prospect of solid dislocations and defects \cite{pollet_2007_superfluidity,boninsegni_2012_colloquium} containing superfluid components that mimic irrotational flow expected in a supersolid.  The debate over $^4\text{He}$ supersolidity highlights the need for an observable to detect simultaneous ODLRO and DLRO.

The search for experimental probes of supersolids has become more pressing with prospects for engineered \cite{scarola_2005_quantum}  supersolidity with ultracold bosonic atoms \cite{lu_2011_strongly,leonard_2017_supersolid,leonard_2017_monitoring,li_2017_stripe,tanzi_2019_observation,bottcher_2019_transient,tanzi_2019_supersolid,chomaz_2019_longlived,guo_2019_lowenergy,norcia_2021_twodimensional,ilzhofer_2021_phase,sohmen_2021_birth,tanzi_2021_evidence,norcia_2022_can,bland_2022_twodimensional,sanchez-baena_2023_heating,chomaz_2022_dipolar}, particularly in highly magnetic atoms such as $^{164}\text{Dy}$.  The tunable dipole-dipole interaction between $^{164}\text{Dy}$ atoms allows strong interaction to drive the formation of high-density quantum droplets \cite{wachtler_2016_quantum,wachtler_2016_groundstate,baillie_2018_droplet,mishra_2020_selfbound} breaking translational invariance \cite{chomaz_2022_dipolar}.
Crystalline (incoherent) quantum droplets give way to supersolid droplets as the strength of the dipolar interaction is lowered to become comparable to kinetic energy.
Experiments identifying and probing small arrays of dysprosium supersolid droplets \cite{tanzi_2019_observation,guo_2019_lowenergy,tanzi_2019_supersolid,bottcher_2019_transient,chomaz_2019_longlived,ilzhofer_2021_phase,norcia_2021_twodimensional,sohmen_2021_birth,tanzi_2021_evidence,norcia_2022_can,bland_2022_twodimensional,sanchez-baena_2023_heating} rely on close support from theory
\cite{roccuzzo_2019_supersolid,blakie_2020_supersolidity,halder_2022_control,halder_2023_twodimensional,smith_2023_supersolidity,ripley_2023_twodimensional,mukherjee_2025_selective,bland_2022_alternatingdomain,halder_2024_induced}.

Current $^{164}\text{Dy}$ experiments observe translational symmetry breaking but are limited to just a few oscillations in the density and must be scaled to observe DLRO.  As system sizes are increased to engineer larger supersolids,  dislocations, defects, small domains, and/or phase separation that mimics supersolid order will arise.  Conventional observables, e.g., density or moment of inertia \cite{tanzi_2021_evidence}, in large $^{164}\text{Dy}$ systems will face ambiguities akin to those arising in $^4\text{He}$ \cite{pollet_2007_superfluidity,sengupta_2005_supersolids,kim_2012_absence,boninsegni_2012_colloquium,chan_2013_overview} because they
are averaged separately and must be combined to infer the coexistence of DLRO and ODLRO.

We propose the use of quantized vortices as scalable supersolid probes \cite{gallemi_2020_quantized,roccuzzo_2020_rotating,ancilotto_2021_vortex,bland_2023_vortices,casotti_2024_observation,ghosh_2024_induced,halder_2025_roadmap}.  Several methods have been used to create vortices in atomic Bose-Einstein condensates (BECs). Rotation by stirring \cite{madison_2000_vortex,abo-shaeer_2001_observation,klaus_2022_observation} was recently used to observe vortices in rotating $^{164}\text{Dy}$ in the supersolid regime \cite{casotti_2024_observation}. Other methods include quenching \cite{sadler_2006_spontaneous,weiler_2008_spontaneous,lamporesi_2013_spontaneous,donadello_2016_creation,goo_2021_defect}, optical phase-engineering and holographic methods \cite{matthews_1999_vortices,denschlag_2000_generating,brachmann_2011_inducing,kumar_2018_producing}, and Laguerre-Gaussian (LG) beams \cite{allen_1992_orbital,marzlin_1997_vortex,andersen_2006_quantized,beattie_2013_persistent,mondal_2014_angular,bhowmik_2016_interaction,das_2020_transfer,ghoshdastidar_2022_pattern}.

Quantized vortex dipoles, readily created in BECs \cite{freilich_2010_realtime,neely_2010_observation,brachmann_2011_inducing,samson_2016_deterministic}, move as quasiparticles making them excellent candidates for recent experimental advances \cite{haljan_2001_use,anderson_2000_vortex,engels_2002_nonequilibrium,freilich_2010_realtime,wilson_2015_situ,serafini_2017_vortex,seo_2017_observation,kwon_2021_sound}.   Example implementations of \emph{in situ} vortex trajectory imaging include an extracting, expanding, and imaging method \cite{freilich_2010_realtime} and high resolution absorption \cite{kwon_2021_sound}.   \emph{In situ} nondestructive imaging of vortex core locations was achieved using dark-field imaging without expanding the BEC \cite{wilson_2015_situ}, and \emph{in situ} vortex dynamics were also observed using atom self-averaging \cite{serafini_2017_vortex}.

We propose vortex-vortex correlations in trajectory imaging as quantitative probes of the fundamental duality inherent to supersolids, Fig. \figref{fig:schema_SS_LG_combo}(b).  Our central observation is that quantized vortices necessitate a superfluid component but also move along density gradient contours \cite{mason_2008_motion,simula_2018_vortex,groszek_2018_motion,fischer_1999_motion} to reveal the underlying lattice structure of a supersolid.
Figure \figref{fig:schema_SS_LG_combo}(c) depicts a proposed experiment wherein a pair of LG beams creates vortex dipoles in a supersolid, and subsequent imaging \cite{wilson_2015_situ,serafini_2017_vortex} records trajectories.  We numerically test the efficacy of our setup.  We consider harmonic traps (containing a few droplets in the supersolid regime) and uniform traps of droplet arrays \cite{gaunt_2013_boseeinstein} more conducive to observing DRLO. We find \cite{supplemental} that our correlation function distinguishes the supersolid from the superfluid and, in the uniform trap, reveals the supersolid lattice structure even in the presence of condensate dissipation \cite{choi_1998_phenomenological,penckwitt_2002_nucleation,proukakis_2008_finitetemperature}. We neglect equilibrium thermal effects since vortex annihilation timescales we find are much shorter than lifetimes of vortices at finite temperature  \cite{rooney_2010_decay}, and, furthermore, finite temperature can even enhance supersolidity \cite{sanchez-baena_2023_heating}.

Our proposal for a unique supersolid order parameter and its experimental implementation establishes a method for identifying long-range order in supersolids, measuring lattice properties, tracking phase transitions, and monitoring supersolid lattice dynamics as experiments with ultracold atoms increase in size.

\emph{Model---}\noindent
We consider $N$ $^{164}$Dy atoms trapped in an external potential $V_{\text{ext}}(\vb{r})$ interacting via \cite{pethick_2008_boseeinstein,pitaevskii_2016_boseeinstein,lahaye_2009_physics}
\begin{align}
    V_{\rm c}(\vb{r}) = g\delta(\vb{r}) \hspace{0.1cm} \rm{and} \hspace{0.1cm} V_{\rm dd}(\vb{r})=\frac{\mu_0 \mu^2}{4\pi} \frac{1-3\cos^2\theta}{r^3}.
\end{align}
The contact interaction strength $g$ is related to the scattering length $a$ by $g = 4\pi\hbar^2a/m$ where $m$ is the mass of the atom and $a$ is the $s$-wave scattering length renormalized by the dipolar interaction \cite{ronen_2006_dipolar,bortolotti_2006_scattering,cinti_2017_superfluid}.
The dipolar interaction $V_{\rm dd}$ varies with the angle $\theta$ between the relative position vector $\vb{r}$ and the polarization direction ($z$ axis) of the dipole.  $\mu_0$ and $\mu$ are the vacuum permeability and the magnetic moment of the atom, respectively.

We model the dynamics of the entire system with the extended Gross-Pitaevskii equation (EGPE) \cite{wachtler_2016_quantum,baillie_2016_selfbound}
\begin{align}
    \im\hbar\pdv{t}\psi(\vb{r}) = \Big[- & \frac{\hbar^2}{2m}\laplacian + V_{\text{ext}}(\vb{r}) + g\abs{\psi(\vb{r})}^2 + \gamma \abs{\psi(\vb{r})}^3  \nonumber \\&
        +\int \dd{\vb{r'}} V_{\rm dd}(\vb{r}-\vb{r'}) \abs{\psi(\vb{r'})}^2 \Big]\psi(\vb{r}). \label{eq:egpe}
\end{align}
Here, $\psi(\vb{r})$ is the system wave function normalized with $\int \dd{\vb{r}} \abs{\psi}^2=N$.
The Lee-Huang-Yang \cite{lee_1957_eigenvalues,lima_2011_quantum,petrov_2015_quantum} correction term contains $\gamma(\epsilon_{\rm dd})=(128\sqrt{\pi}\hbar^2a^{5/2}/3m)(1+\frac{3}{2}\epsilon_{\rm dd}^2)$ \cite{schutzhold_2006_meanfield,lima_2012_meanfield}, which depends on the ratio of the dipolar length scale $a_{\rm dd}$ to $a$, $\epsilon_{\rm dd} =a_{\rm dd}/a$, where $a_{\rm dd}=\mu_0\mu^2m/(12\pi\hbar^2)$.
$\epsilon_{\rm dd}$ can be tuned using a Feshbach resonance in $^{164}$Dy to vary $a$ \cite{tang_2018_tuning,chomaz_2022_dipolar}, thus, allowing exploration of phase transitions \cite{baillie_2016_selfbound}.  For $^{164}$Dy, the dipole moment $\mu{=}9.93\mu_B$, sets a dipolar length scale $a_{\rm dd}{=}131 a_0$ where $\mu_B$ and $a_0$ are the Bohr magneton and Bohr radius, respectively.
We choose $a=94a_0 (\epsilon_{\rm dd}=1.39)$ and $110a_0 (\epsilon_{\rm dd}=1.19)$ to achieve the supersolid and superfluid phases, respectively \cite{bottcher_2019_transient}.

To study vortex dynamics in superfluid and supersolid regimes, we assume two different external trapping potentials: (i) uniform in the $xy$ plane and harmonic along the $z$ axis, $V_{\rm ext}{=} m \omega_z^2 z^2/2$ and (ii) an oblate 3D harmonic trap, $V_{\rm ext}{=} m \omega^2(x^2+y^2+\lambda^2 z^2)/2$ where trapping anisotropy $\lambda$ is the ratio of trapping frequency along the $z$ axis to that in the $xy$ plane.  In both cases, we induce vortex dipoles using the phase-imprinting method \cite{leanhardt_2002_imprinting,bandyopadhyay_2017_dynamics} by multiplying the wave function $\psi(\vb{r})$ with a phase factor of $\exp[\im l \tan^{-1}(y-y_0)/(x-x_0) ]$.  Here $l \hbar$, with $l=\pm 1$, is the vortex angular momentum and $(x_0,y_0)$ is the vortex initial position.

\emph{Vortex motion in a uniform trap---}\noindent
First, we consider a uniform planar trap to ignore the effect of density variation due to trapping confinement.  We solve the EGPE \eqref{eq:egpe} using the split-step Crank-Nicolson method \cite{muruganandam_2009_fortran} and study the vortex dynamics in the system by tracking the vortex trajectories over time. In the uniform trap, $N{=}1.3\times10^6$ atoms are trapped along the $z$ axis with $\omega_z{=}2\pi\times 150$ Hz, and hard walls in the $xy$ plane. The simulation box size is $L_x{=}L_y{=}2L_z{=}32.8\,\mu$m and the grid size is $(256\times256\times128)$.
By solving the EGPE, we obtain the superfluid and supersolid ground state for different $\epsilon_{\rm dd}$.
In a uniform trap, the superfluid phase has a flat density profile, whereas the supersolid phase shows a density modulation.  The supersolid crystalline structure of quantum droplets contains a low density background, Fig. \figref{fig:schema_SS_LG_combo}(c).

Vortices follow qualitatively distinct trajectories in the supersolid and superfluid phases.  To see this, we dynamically imprint a vortex dipole. The vortex dipole will have a linear momentum in the superfluid phase which is inversely proportional to the separation $d$ of $l=+1$ and $l=-1$ vortices \cite{sasaki_2010_benard,aioi_2011_controlled}. Therefore, both vortices move in straight lines parallel to each other with a constant velocity in the superfluid due to the Magnus effect \cite{griffin_2020_magnusforce}. In contrast, in the supersolid phase, the vortex trajectories depend on nonuniform and periodic density patterns. In the frame of reference of the vortex, density changes with time and space leading to a nonuniform force on the vortices. Because of this nonuniform force, vortices move in nonlinear contours outlining the supersolid density gradient.  In contrast to the superfluid phase, the vortex dipole is not stable in the supersolid phase and eventually annihilates when the $l=+1$ and $l=-1$ vortices approach each other.  We test these qualitative expectations quantitatively using the EGPE.

\begin{figure}[htbp]
    \centering
    \includegraphics[width=0.49\textwidth]{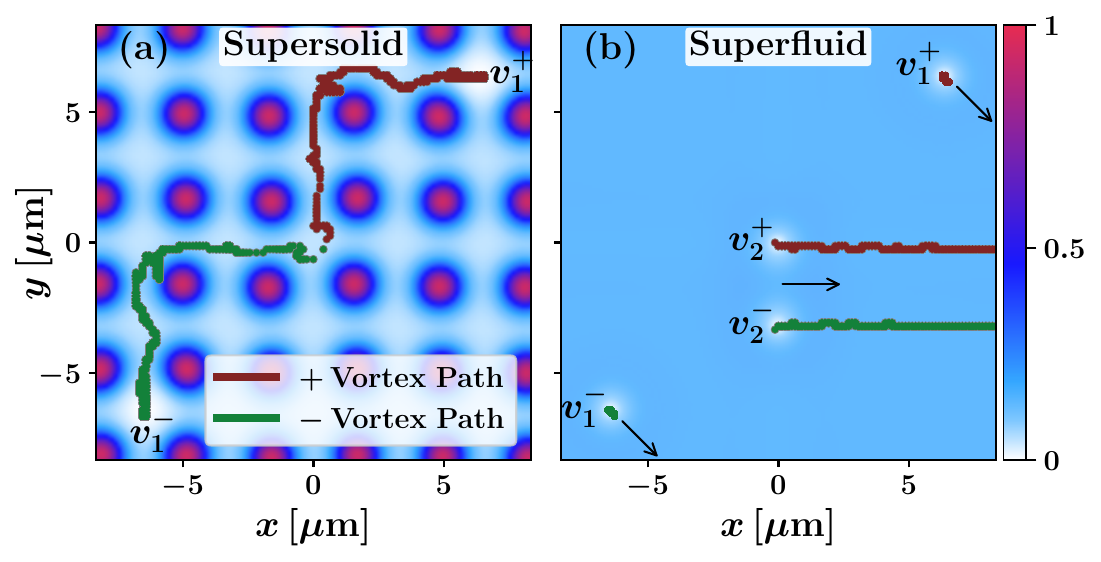}
    \caption{Ground state density profile in the $z=0$ plane and vortex motion in a uniform trap in (a) supersolid and (b) superfluid phases. $v_{\it{i}}^+$ and $v_{\it{i}}^-$ are the initial position of the $l=+1$ and $l=-1$ vortex of $i$ th vortex dipole. The red (green) line represents the trajectory of a $l=+1$ ($l=-1$) vortex. The black arrows in (b) represent the direction of motion of vortex dipoles in the superfluid phase. In (b), the simulations of vortex dipoles $(v_1^+, v_1^-)$ and $(v_2^+, v_2^-)$ are done separately and the background density is the product of two ground states to identify the vortices.
    }
    \label{fig:vortex_motion}
\end{figure}

Figure \figref{fig:vortex_motion} shows the vortex trajectories in the superfluid and supersolid phases.
We choose a vortex dipole $(v_1^+, v_1^-)$ in both phases as shown in Figs. \figref{fig:vortex_motion}(a) and \figref{fig:vortex_motion}(b). The $l=+1$ and $l=-1$ vortices follow nonuniform trajectories and eventually annihilate in the supersolid phase, while in the superfluid phase, they move in straight lines parallel to each other. The vortex dipole motion in the superfluid phase is stable, and they do not annihilate.

We verify that the velocity of the vortices is inversely proportional to the separation.  In the superfluid phase, we prepare a well separated $l=\pm 1$ vortex pair $(v_1^+, v_1^-)$.  Figure \figref{fig:vortex_motion}(b) shows that the distances traveled by this vortex dipole are small. In Fig. \figref{fig:vortex_motion}(b), we also show motion of a closely spaced vortex dipole $(v_2^+, v_2^-)$ in the superfluid phase [this is studied separately from $(v_1^+, v_1^-)$, not shown in Fig. \figref{fig:vortex_motion}(a) for clarity of the figure]. These vortices traveled a longer distance (full paths are not shown) compared to the vortex dipole $(v_1^+, v_1^-)$.

\emph{Correlation of vortex trajectories---}\noindent
Now, we turn to our central proposal.  Given the distinct behavior of vortex motion in each phase, we seek to extract quantitative information from the vortex trajectories.  We define a vortex-vortex trajectory correlation function
\begin{align}
    C(\vb{r}) = \expval{s(\vb{r}_0)s(\vb{r}_0+\vb{r})},
    \label{eq:correlation}
\end{align}
where $s(\vb{r})$ is 1 if a vortex of any $l$ passes through $\vb{r}$ and 0, otherwise.  In the following calculations, we choose to compute $C(\vb{r})$ using trajectories of opposite $l$ with no loss of generality.  Note that $C(\vb{r})$ vanishes in the quantum droplet phase (because there is no phase coherence for vortices to form) and should show only trivial structure in the superfluid phase because the vortices move in straight lines.
But in the supersolid phase, measurements of $C(\vb{r})$ will probe the underlying crystal structure of the supersolid.
Therefore, $C(\vb{r})$ can be used as a supersolid order parameter.  Importantly, the initial configuration of vortices must be chosen to break the underlying translational symmetry of the supersolid phase, otherwise, straight-line trajectories will arise \cite{supplemental}.  Straight trajectories will not allow measurements to distinguish between superfluid and supersolid states.  In the following, we choose the initial positions of the vortex dipoles on the opposite arcs of a circle in the simulations to avoid biased choices in probing the quantum droplet square lattice in the supersolid phase.

\begin{figure}[b]
    \centering
    \includegraphics[width=0.49\textwidth]{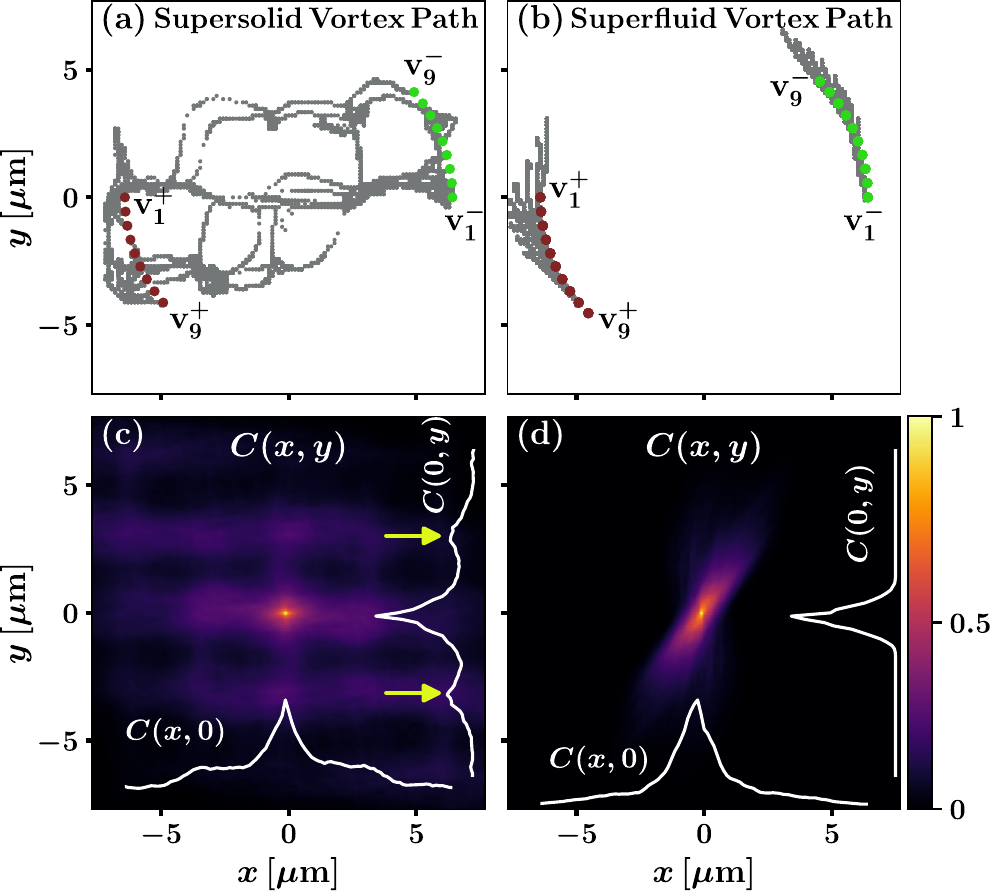}
    \caption{Combined vortex trajectories of nine vortex dipoles $[(v_1^+, v_1^-),\dots,(v_9^+, v_9^-)]$ from separate simulations in (a) supersolid and (b) superfluid phases. The red (green) dots represent the initial position of the $l=+1$ ($l=-1$) vortex of the vortex dipoles. The grey lines represent the vortex trajectories. The correlation function $C(\vb{r})$ of the vortex trajectories in (c) superfluid and (d) supersolid phases. (c) and (d) show the correlation function $C(\vb{r})$ of the vortex trajectories in the supersolid and superfluid phases, respectively. In (c), apart from the primary peak at $\vb{r}=0$, there are the secondary peaks at $\vb{r}\neq0$ denoted by yellow arrows for $C(0,y)$. In contrast, in (d), the correlation function $C(\vb{r})$ shows only the primary peak at $\vb{r}=0$. All simulation parameters are the same as in Fig. \figref{fig:vortex_motion}.}
    \label{fig:vortex_corr}
\end{figure}

We test the utility of $C(\vb{r})$ using EGPE simulations designed to replicate repeated measurements in a uniform trap.  Figures \figref{fig:vortex_corr}(a) and \figref{fig:vortex_corr}(b)
show nine separate simulations of the resulting motion of the vortex dipoles $[(v_1^+, v_1^-),\dots,(v_9^+, v_9^-)]$ in both superfluid and supersolid phases. For all the vortex dipoles, the $l=+1$ and $l=-1$ vortex separation $d\approx 13{\,\mu\rm m}$ is the same as shown by the initial position of the $l=+1$ ($l=-1$) vortex by red (green) dots in Figs. \figref{fig:vortex_corr}(a) and \figref{fig:vortex_corr}(b).

We locate the vortex position at different instants of time and find the trajectories of the vortices as time evolves. We find different trajectories for different sets of initial positions of the vortices. By merging all of the trajectories we obtain a combined 2D trajectory on the $xy$ plane shown in Figs. \figref{fig:vortex_corr}(a) and \figref{fig:vortex_corr}(b).

We find that even with a limited number of trajectories, $C(\vb{r})$ reveals aspects of the underlying lattice structure of the supersolid.  Figure \figref{fig:vortex_corr}(c) shows that the correlation function $C(\vb{r})$, apart from the primary trivial peak at $\vb{r}=0$, shows secondary peaks at $\vb{r}\neq0$ due to the nonuniform and periodic density patterns in the supersolid phase, whereas in the superfluid phase, $C(\vb{r})$ shows only the primary peak at $\vb{r}=0$ due to the straight-line motion of the vortices. Thus, the correlation function $C(\vb{r})$ of vortex trajectories is a useful tool for distinguishing the superfluid and supersolid phases.

Use of $C(\vb{r})$ to probe a supersolid requires the following conditions to be met:  (i) As mentioned above, initial positions must break the supersolid crystal symmetry.  If initial vortex positions are chosen, for example, to be random, a sufficient number of trajectories must be imaged to resolve auxiliary peaks.  The limited number of runs in Fig. \figref{fig:vortex_corr} were chosen near this threshold,  (ii) low dissipation or heating,  (iii) vortices are, ideally, to be used as noninvasive probes that leave the supersolid intact.  We assume an LG beam spot size ($\sim 1 \,\mu\text{m}$) below the supersolid lattice spacing ($\sim 3 \,\mu\text{m}$ in our simulations) and with initial intervortex spacing high enough to avoid high vortex speeds.  (iv) We assume the use of a uniform trap (or a sufficiently large number of atoms) to allow resolution of translational symmetry breaking and DLRO. Supplementary material \cite{supplemental} shows data relaxing conditions (i) and (ii) in order to map the supersolid lattice whereas the following simulations relax conditions (iii) and (iv).

\emph{Vortex motion in a harmonic trap---}\noindent
Now, we turn to small system sizes and harmonic trapping relevant for ongoing experiments.
Tight trapping will limit the range of the solid and can even lead to spurious supersolid signatures \cite{scarola_2006_searching}.  We choose $N{=}80\,000$ atoms to be trapped in an oblate-shaped 3D harmonic trap with trapping frequencies $(\omega,\omega_z){=}2\pi\times (45,133)$ Hz \cite{halder_2022_control}, where $\omega$ is the transverse trapping frequency. The simulation box size is $L_x{=}L_y{=}L_z{=}30\,\mu$m.

We studied the motion of vortex dipoles in both superfluid ($a=110 a_0$) and supersolid ($a=94 a_0$) regimes.  Without any vortices, the density profile of the superfluid is Gaussian, whereas the supersolid regime shows a single quantum droplet in the center of the trap with a low-density superfluid background \cite{bottcher_2019_transient,chomaz_2019_longlived}.
In both regimes, we prepare ground states with a vortex dipole $(v_1^+, v_1^-)$ separated by a distance $d\approx 6\,\mu\rm m$, and we observe the motion of the vortices by tracking the vortex trajectories over time.

\begin{figure}[t]
    \centering
    \includegraphics[width=0.48\textwidth]{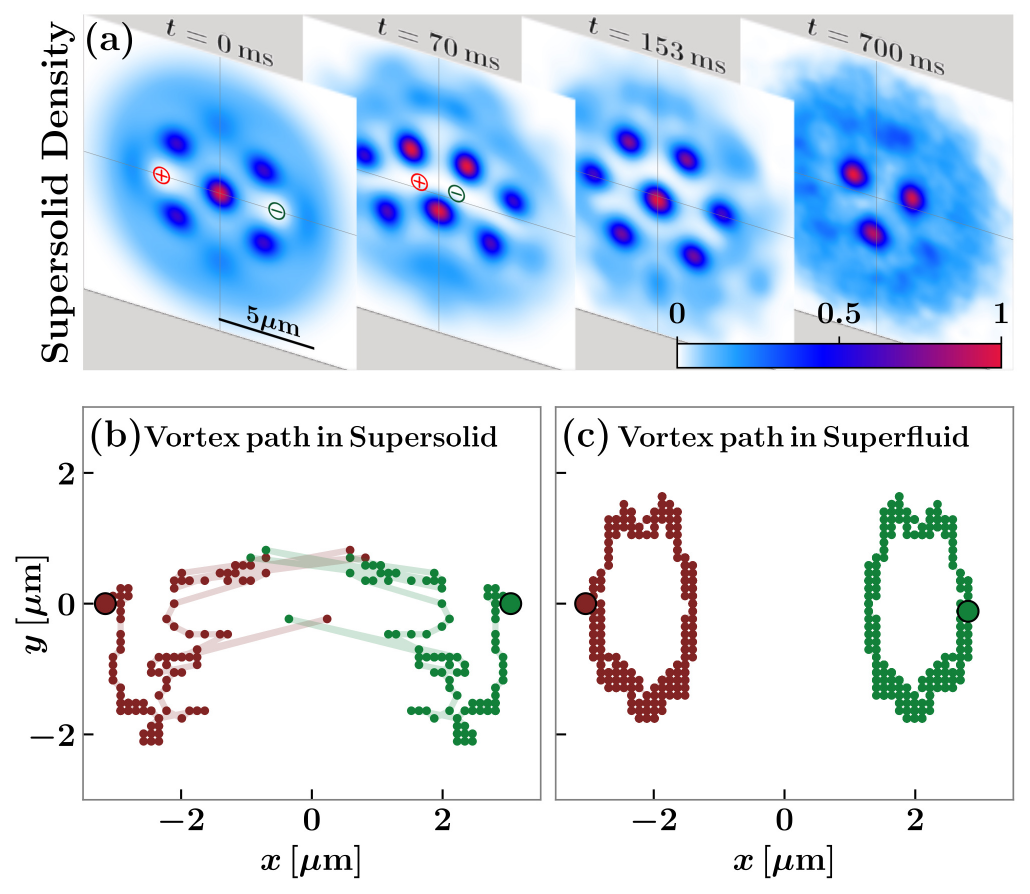}
    \caption{(a) Snapshots of density profiles in the $z=0$ plane at different times in the harmonic trap after the vortex dipole is imprinted in the supersolid regime. The red (green) circled ``+''(``-'') marker represents the $l=+1$ ($l=-1$) vortex. (b) and (c) show the $l=+1$ (red dots) and $l=-1$ (green dots) vortex trajectories in the supersolid and superfluid regimes, respectively.
    }
    \label{fig:vortex_motion_harmonic}
\end{figure}

Figure \figref{fig:vortex_motion_harmonic}(a) shows density profiles in the supersolid regime after imprinting the vortex dipole. At $t=0$ the creation of the vortices causes the single droplet at the center to split into multiple droplets. Over time, these droplets deform and form continuously at different locations as a result of the motion of the vortices.  Here, the dynamics explores a variety of metastable configurations in the supersolid regime.  Once the $l=\pm1$ vortices annihilate each other, an equilibrium state with three droplets is achieved, as we see in $t=700\,\rm ms$ in \figref{fig:vortex_motion_harmonic}(a). The trajectories of the vortices are affected by the density modulation in the supersolid regime, moving in nonlinear paths before annihilation [Fig. \figref{fig:vortex_motion_harmonic}(b)]. In contrast, the vortices in the superfluid regime do not annihilate.  They keep moving in an elliptical path with a constant speed, as shown in Fig. \figref{fig:vortex_motion_harmonic}(c) \cite{li_2008_vortex,middelkamp_2011_guidingcenter}.  Therefore, we see that vortex dipole trajectories in the supersolid and superfluid regimes show qualitatively distinct behavior even in small systems with tight trapping.

\emph{Discussion---}\noindent
We introduced and numerically tested a supersolid order parameter defined by correlations in vortex trajectories.  Our tests used a limited number of trajectories to replicate sampling for near-term experiments.  Larger numbers of samples allow resolution of complete supersolid lattice structures, relevant for larger system sizes \cite{supplemental}.  Our correlation function can also be used to study important properties of  supersolids, including lattice dynamics due to supersolid phonons and phase transitions.  Possible applications to different systems include exploration of broken translational symmetry in thin films of superfluid $^{4}\text{He}$ \cite{nakamura_2016_possible,nyeki_2017_intertwined,choi_2021_spatially} and Fulde-Ferrell-Larkin-Ovchinnikov \cite{fulde_1964_superconductivity,larkin_1964_nonuniform} superconductors.

\emph{Acknowledgments---}\noindent
S.D. thanks Soumyadeep Halder and Bankim C. Das for the useful discussions. We acknowledge support from Grants No. AFOSR-FA9550-23-1-0034 and No. FA9550-19-1-0272, and from ARO Grant No. ARO-W911NF2210247.

\bibliography{references}
\cleardoublepage
\end{document}


\title{Supplemental Material for ``Unveiling Supersolid Order via Vortex Trajectory Correlations''}

\author{Subrata Das}
\author{Vito W. Scarola}
\affiliation{Department of Physics, Virginia Tech, Blacksburg, Virginia 24061, USA}

\maketitle

\section{Square Lattice Vortex Trajectory Correlation Functions }

In the main text, we have shown the vortex trajectories for the vortex dipole where the $l=+1$ and $l=-1$ vortices are located on opposite arcs of a circle for the case of a uniform trap. Those trajectories break lattice symmetries and, therefore, will distinguish between the supersolid and superfluid regimes, whether there is a square or triangular lattice pattern in the supersolid. Furthermore, such symmetry breaking sampling of vortex trajectories will also yield the lattice spacing.  Here, in the example of a square lattice supersolid, we assume prior knowledge of the lattice spacing and an approximate knowledge of the lattice structure.  We demonstrate the ability of post-selection of the vortex trajectory correlations to yield more quantitative information on supersolid lattice structure.

\begin{figure}[htbp]
    \centering
    \includegraphics[width=\textwidth]{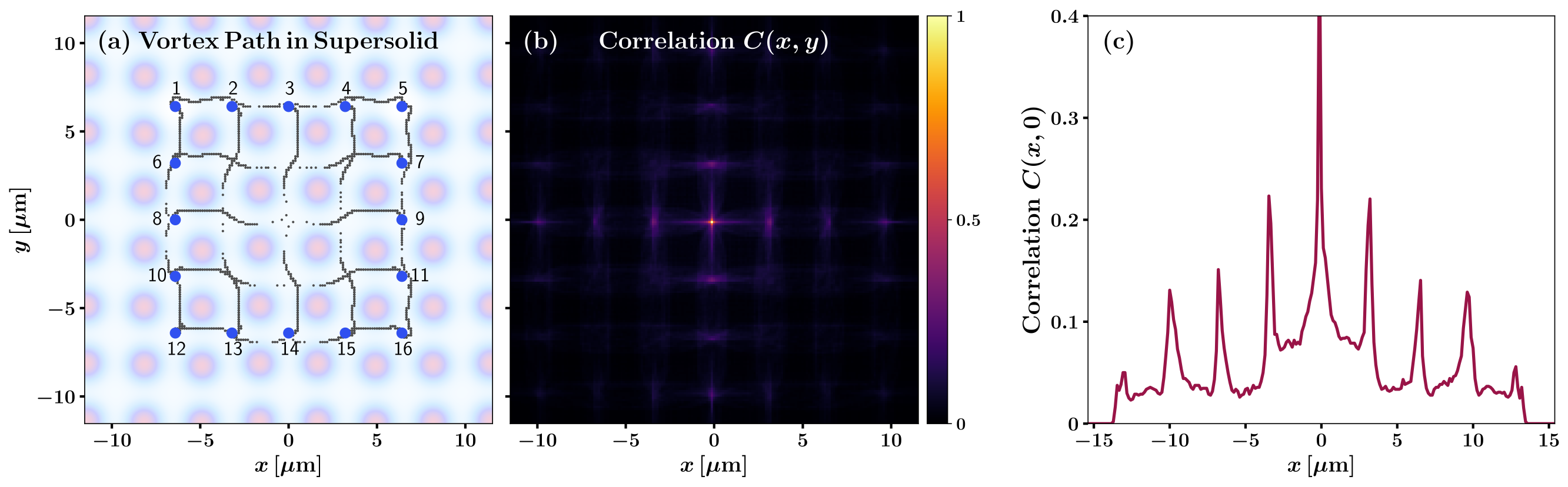}
    \caption{Vortex trajectories and their correlations. (a) Vortex trajectories for different locations of the vortex dipole $(v^+,v^-)$ from different simulations, where $(v^+,v^-)\equiv\{(1,12),(2,13),(3,14),(4,15),(5,16),(1,5),(6,7),(8,9),(10,11),(12,16)\}$ and the indices (1-16) correspond to the $l=+1$ or $l=-1$ vortex locations marked with blue marker. Trajectory of the vortex dipole $(v^+,v^-)$ is an approximately straight line connecting $v^+$ and $v^-$. The opaque background shows the supersolid density of the ground state without vortices. Parameters, except for the initial vortex locations, are the same as Fig.~3a of the main text. (b) Vortex-vortex correlation function $C(x,y)$ calculated from different trajectories.  (c) The correlation $C(x,0)$ along $x$-axis.}
    \label{fig:supple_vortex_tra_corr}
\end{figure}

The following simulations are performed to replicate a proposed experimental sampling protocol given lattice spacing and a square lattice hypothesis.  We assume a large collection of data where initial vortex dipoles are imprinted to map out the lattice.  Consider the set of data where vortices are imprinted near density maxima.  For example,  imagine a vortex dipole whose $l=+1$ vortex is located between 1 and 2 in Fig.~\figref{fig:supple_vortex_tra_corr}(a), and the $l=-1$ vortex is located between 12 and 13. In that case, their trajectory will follow a \emph{curved} path.  We remove such curved trajectories from our data set.

We now consider trajectories to keep in the dataset.  $l=+1$ and $l=-1$ vortices placed at density minima between droplets will move in nearly straight lines toward each other.  In Fig. \figref{fig:supple_vortex_tra_corr}(a), if we choose the vortex dipole $(v^+,v^-)\equiv(1,12)$, their trajectory is a nearly straight line connecting the $l=+1$ vortex (point 1) and $l=-1$ vortex (point 12).  By post-selecting quasi-straight line trajectories we can obtain a
vortex correlation function map of the supersolid lattice.  Note that the vortex placement and post-selection process described here are biased and assume forehand knowledge of the lattice spacing and structure obtained from the method discussed in the main text.

Once quasi-straight line vortex trajectories have been post-selected, we can build the correlation function.
We obtain the dynamics of the vortex dipole for different locations of the vortex dipole. 
All vortex dipoles are studied separately with the same parameters used in the main text for the uniform trap. Using these trajectories, we calculate the vortex-vortex correlation function $C(x,y)$ [Fig. \figref{fig:supple_vortex_tra_corr}(b)] and the correlation $C(x,0)$[Fig. \figref{fig:supple_vortex_tra_corr}(c)] along the $x$-axis. From the secondary peaks of the correlation function $C(x,0)$, we can identify and find the periodicity of the square lattice structure.

\section{Dissipation Effects}

\begin{figure}[htbp]
    \centering
    \subfloat[][]{\includegraphics[width=0.4\textwidth]{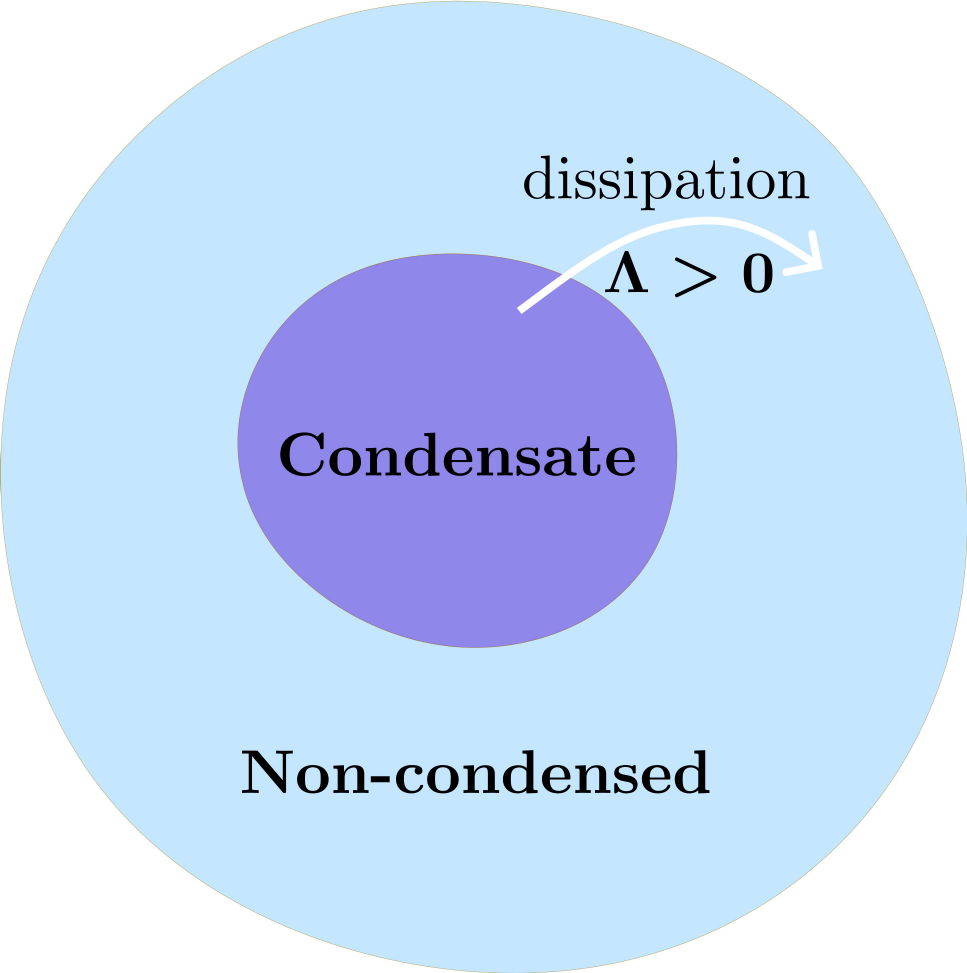}}\qquad \qquad
    \subfloat[][]{\includegraphics[width=0.45\textwidth]{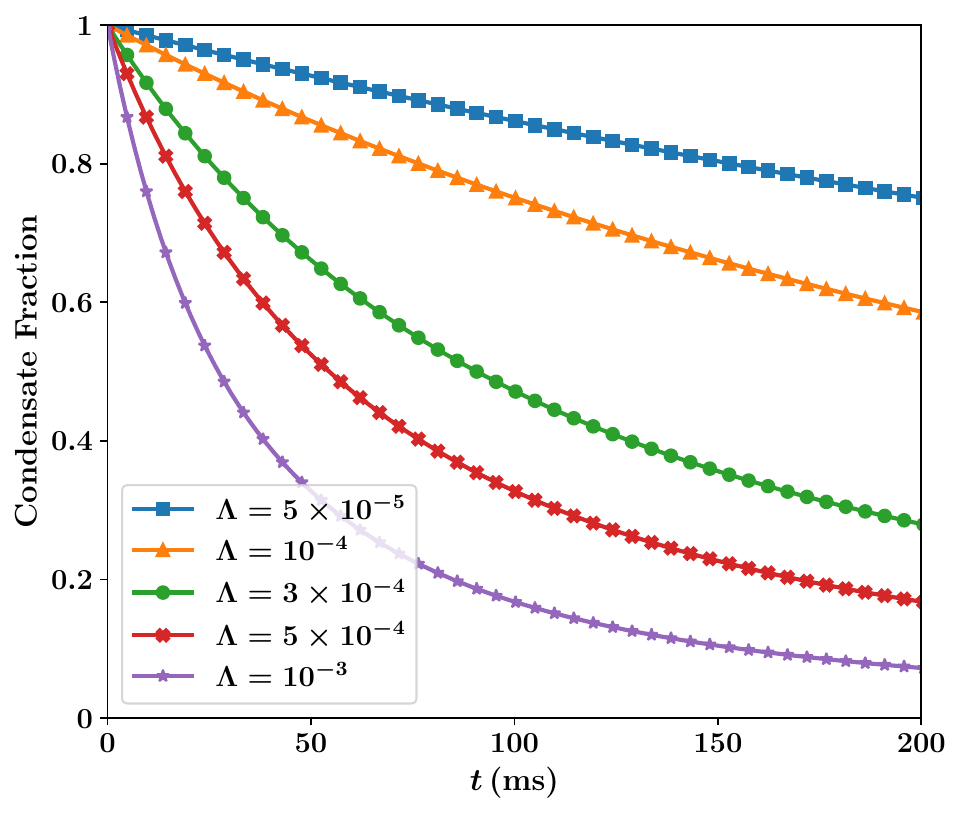}}
    \caption{(a) Schematic of atom loss due to dissipation. (b) Condensate fraction over time for different damping factors, $\Lambda$. All other parameters are the same as in Fig. 3 in the main text.}
    \label{fig:condensate_thermal}
\end{figure}

In this section, we quantify the limits of our proposed measure in the presence of dissipation.  We assume that the supersolid is first prepared at zero temperature.  We then model perturbations and heating by turning on a phenomenological dissipation parameter that leads to the loss of atoms from the condensed part of the wavefunction, Fig.~\ref{fig:condensate_thermal}(a).  Dissipation has been connected to a quasi-equilibrium temperature in quantum gases containing vortices \cite{choi_1998_phenomenological,penckwitt_2002_nucleation}.  In this phenomenological picture the dissipative term models the loss of atoms from the condensate due to collisions between condensed and non-condensed atoms and therefore approximates heating.

We find that dissipation creates two different regimes depending on its strength.  For weak dissipation, vortex trajectories are shortened, thus lowering signal-to-noise in our proposed observable, but correlations are otherwise preserved.  Here the losses can be roughly thought of as slow heating as the condensate fraction decreases while the supersolid survives.  High dissipation, by contrast, quickly destroys the supersolid phase and, in turn, significantly impacts vortex trajectories.  Here the condensate fraction depletes too quickly for appreciable information to be extracted from vortex trajectories.

To include dissipation, we consider the extended Gross-Pitaevskii equation with a phenomenological damping factor $\Lambda$ \cite{proukakis_2008_finitetemperature}:
\begin{align}
    i\hbar\pdv{t}\psi(\vb{r}) = (1-i\Lambda)\Big[- & \frac{\hbar^2}{2m}\laplacian + V_{\text{ext}}(\vb{r}) + g\abs{\psi(\vb{r})}^2 + \gamma \abs{\psi(\vb{r})}^3
        +\int \dd{\vb{r'}} V_{\rm dd}(\vb{r}-\vb{r'}) \abs{\psi(\vb{r'})}^2 \Big]\psi(\vb{r}).  \label{eq:egpe_damped}
\end{align}
$\Lambda > 0$ controls the rate of atom loss of the condensate to the thermal cloud due to collisions \cite{choi_1998_phenomenological} as shown in Fig. \figref{fig:condensate_thermal}(a).  We calculate the condensate fraction for different damping factors $\Lambda$, as shown in Fig. \figref{fig:condensate_thermal}(b). The condensate fraction is the ratio of the number of atoms in the condensate at time $t$ to the initial number. For smaller damping factors (between $\Lambda=5\times10^{-5}$ and $1\times10^{-4}$), the decay of the condensate fraction is almost linear on time scales of interest, while for larger damping factors (from $\Lambda=5\times10^{-4}$ to $10^{-3}$), the decay is exponential.

\par
\begin{figure}[htbp]
    \centering
    \includegraphics[width=\textwidth]{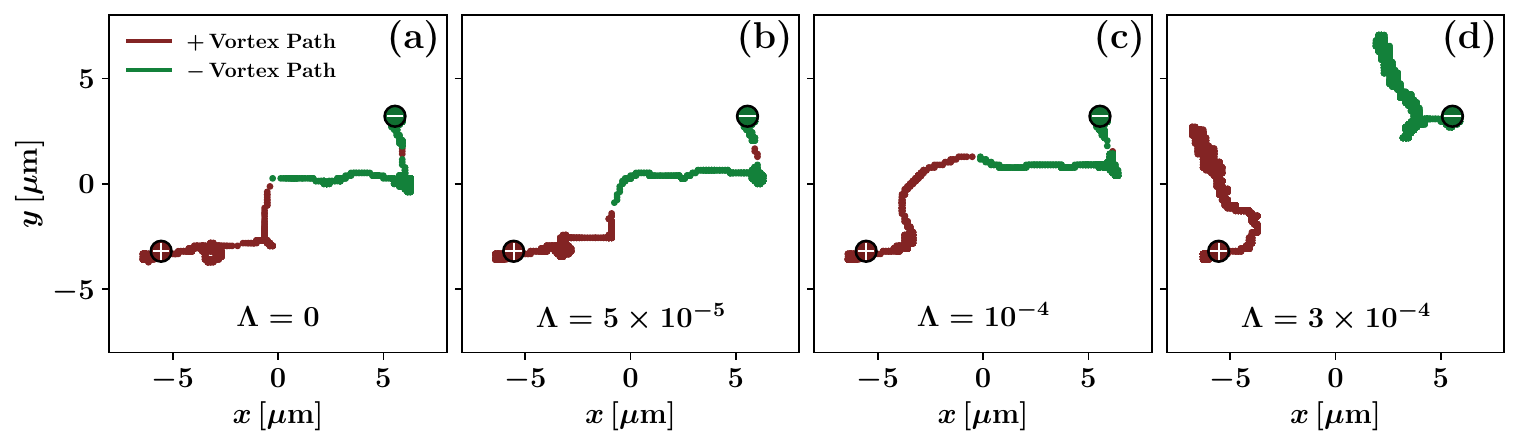}
    \caption{Vortex trajectories for different damping factors $\Lambda$. For damping factors $\Lambda=5\times10^{-5}$ and $10^{-4}$, the vortex-antivortex annihilation dynamics occur. For a larger damping factor ($\Lambda=3\times10^{-3}$), vortex trajectories are nearly linear, as in the superfluid. All simulation parameters are the same as those used for the $\Lambda=0$ limit in Fig. 3 in the main text.}
    \label{fig:supple_vortex_tra_damp}
\end{figure}

A larger $\Lambda$ results in a quicker reduction of supersolidity, while a smaller $\Lambda$ allows the condensate to retain its supersolid nature for a longer period, enabling observation of vortex-antivortex annihilation dynamics. The vortex trajectories shown in Figs. \figref{fig:supple_vortex_tra_damp}(a)-\figref{fig:supple_vortex_tra_damp}(c), demonstrate vortex-antivortex annihilation for damping factors ($\Lambda=0-10^{-4}$). In contrast, for a larger damping factors (\emph{e.g.}, $\Lambda=3\times10^{-4}$) in Fig. \figref{fig:supple_vortex_tra_damp}(d) the vortex and antivortex no longer annihilate. Instead they move in the same direction, exhibiting trajectories typical of superfluid behavior. Notably, for larger damping factors, such as $\Lambda=10^{-3}$, vortex-antivortex annihilation can still occur if their initial positions are close enough, with the distance approximately scaling as $\Lambda^{-1}$. \par
\begin{figure}[H]
    \centering
    \includegraphics[width=\textwidth]{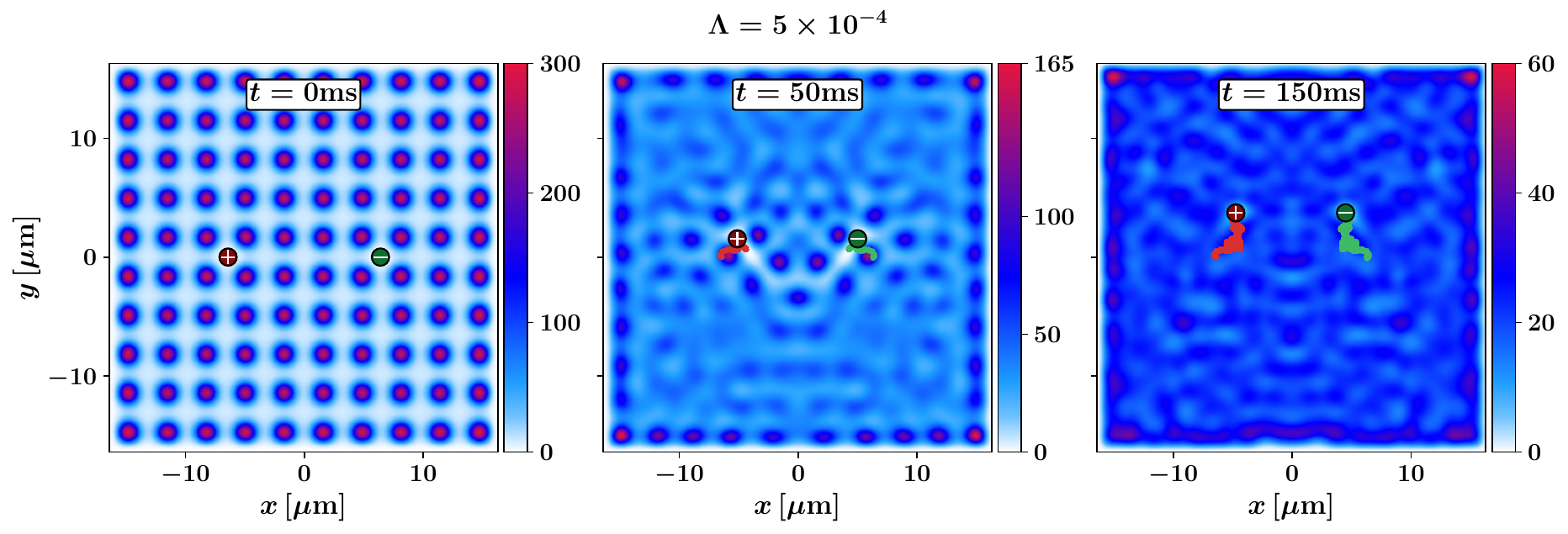}
    \caption{Density profiles and vortex trajectories at times $t=0,50$, and 150 ms for damping factor $\Lambda=5\times 10^{-3}$. All other parameters are the same as in Fig. 3 in the main text.  The rightmost panel shows that here, the loss of coherence and supersolidity due to dissipation coincides with the slowing of the vortices and the absence of correlations.    }
    \label{fig:supple_den_vortex_damp}
\end{figure}
To understand the impact of the damping factor on the density profile and vortex trajectories when vortex-antivortex pairs do not annihilate, we set $\Lambda=5\times10^{-4}$ as shown in Fig. \figref{fig:supple_den_vortex_damp}. When there is a high enough condensate fraction to show supersolidity, the vortices experience non-uniform forces due to the density gradients and move toward each other. However, as the condensate depletes over time, the density gradient decreases, causing the system to lose its supersolid properties (\emph{e.g.}, at $t=50\rm\,ms$). Therefore, the vortex and antivortex no longer move toward each other; instead, they move in the same direction, similar to their behavior in a superfluid as we see at $t=150\rm\, ms$.  These nearly parallel trajectories are randomized by the noisy background caused by dissipation.  As an empirical rule of thumb, we find that once dissipation depletes the condensate by $\sim 50\%$, supersolid correlations cannot be measured due to the loss of density gradients imposed by the solid and slowing of the vortices.

\begin{figure}[htbp]
    \centering
    \includegraphics[width=\textwidth]{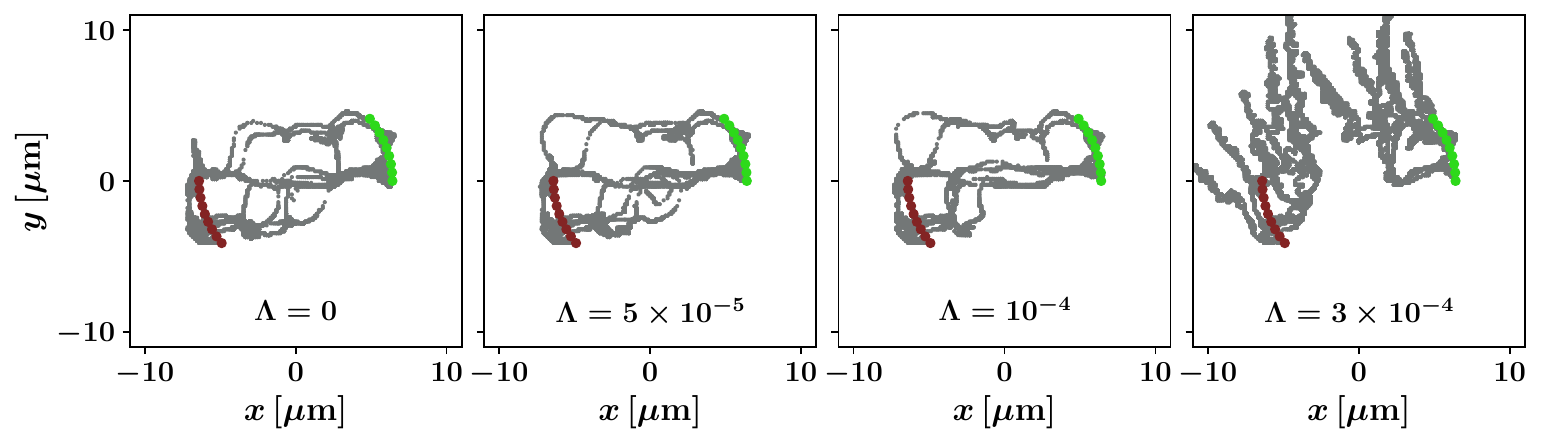}
    \caption{Combined vortex trajectories from different initial vortex positions for different damping factors. All other parameters are same as in Fig. 3 in the main text. }
    \label{fig:supple_vor_traj_damp_all}
\end{figure}

\begin{figure}[htbp]
    \centering
    \subfloat[]{\includegraphics[width=0.45\textwidth]{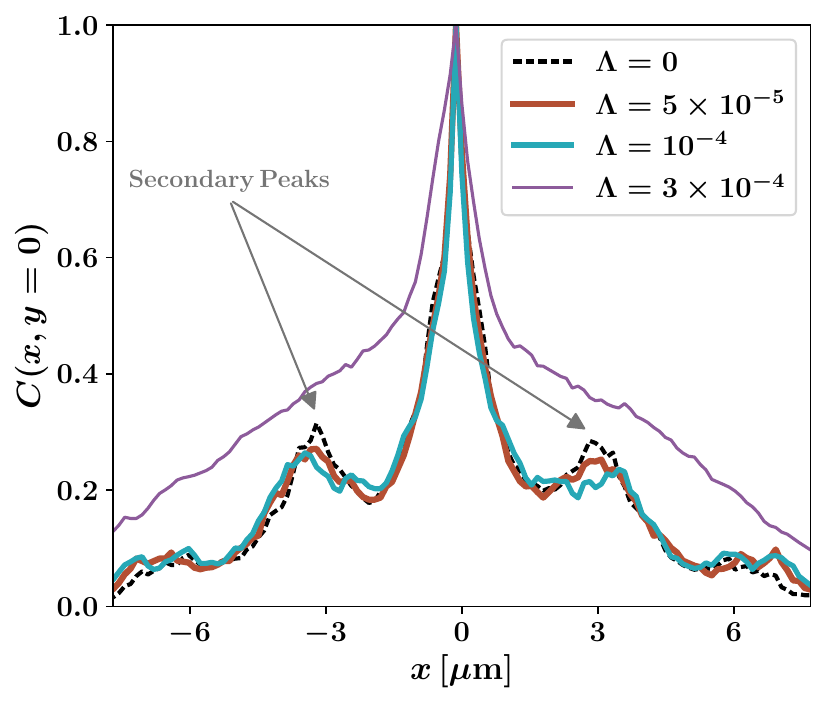}}
    \subfloat[]{\includegraphics[width=0.45\textwidth]{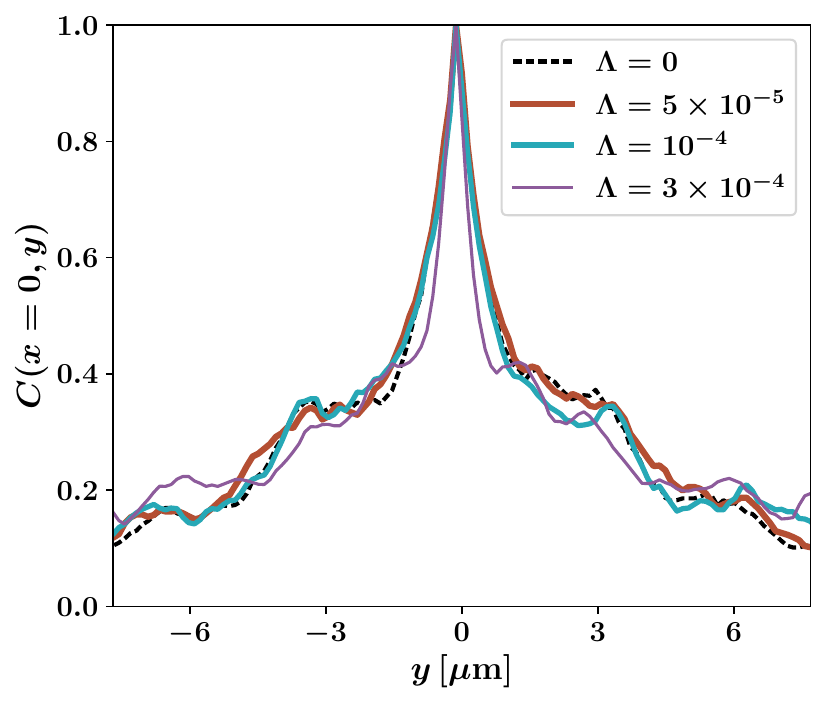}}
    \caption{Correlation function $C$ of vortex-vortex trajectories along the (a) $x$-axis and (b) $y$-axis for different damping factors $\Lambda$. All other parameters are the same as in Fig. 3 in the main text.  Here we see that high dissipation erodes the secondary peaks, concomitant with the loss of supersolidity. }
    \label{fig:supple_vortex_corr_damp}
\end{figure}

For nonzero damping factors, when the vortex-antivortex annihilation dynamics occur, we calculate the correlation of the trajectories for different configurations as we did in the main text (Fig.~3).  Fig.~\figref{fig:supple_vor_traj_damp_all} shows a collection of vortex trajectories taken from several different initial configurations.  Using these trajectories we obtain correlation functions $C(x,0)$ along the $x$-axis and $C(0,y)$ along the $y$-axis as shown in Figs. \figref{fig:supple_vortex_corr_damp}(a) and \figref{fig:supple_vortex_corr_damp}(b), respectively.
Low $\Lambda$ data show behavior similar to the $\Lambda=0$ case discussed in the main text.  The values of secondary peaks are slightly smaller than the $\Lambda=0$ case.  Nonetheless, the periodicity of the lattice structure can be identified from the peaks.  The high $\Lambda$ data show a complete loss of structure, consistent with the absence of supersolidity.

The numerical results presented in this section reveal that as long as the condensate fraction is kept above $\sim 50\%$, supersolid correlations are still visible in our proposed vortex-vortex correlation function despite dissipation.

\bibliography{references}